\documentclass[11pt,a4paper]{article}
\pdfoutput=1
\usepackage{jheppub1}
\usepackage{amsfonts,amsmath,amssymb}
\usepackage{enumerate}
\usepackage{hyperref}
\usepackage{graphicx}

\newcommand{\be}{\begin{equation}}
\newcommand{\ee}{\end{equation}}

\newcommand{\bea}{\begin{eqnarray}}
\newcommand{\eea}{\end{eqnarray}}

\newcommand{\bes}{\begin{subequations}}
\newcommand{\ees}{\end{subequations}}

\newcommand{\nn}{\nonumber}

\title{Comments on the symmetry of AdS$_6$ solutions in String/M-theory and Killing spinor equations}
\author{Hyojoong Kim}
\author{and Nakwoo Kim}
\affiliation{Department of Physics 
and Research Institute of Basic Science, \\ Kyung Hee University, 
Seoul 02447, Korea}
\emailAdd{h.kim@khu.ac.kr}
\emailAdd{nkim@khu.ac.kr}

\abstract{
It was recently pointed out in \cite{Kim:2015hya} that AdS$_6$ solutions in IIB theory enjoy an extended 
symmetry structure and the consistent truncation to $D=4$ internal space leads to a nonlinear sigma model with target $SL(3,\mathbb{R})/SO(2,1)$. We continue to study the purely bosonic $D=4$ effective action, and elucidate how the addition of scalar potential term still allows Killing spinor equations in the absence of gauge fields. 
In particular, the potential turns out to be a single diagonal component of the coset representative.
Furthermore, we perform a general analysis of the integrability conditions of Killing spinor equations 
and establish that the effective action can be in fact generalized
to arbitrary sizes and signatures, e.g. with target $SL(n,\mathbb{R})/SO(p,n-p)$ and the scalar
potential expressible by a single diagonal component of the coset representative. We also comment on 
a similar construction and its generalizations 
of effective $D=5$ purely bosonic non-linear sigma model action related to AdS$_6$ in M-theory. 
}

\begin{document}
\maketitle
\flushbottom


\section{Introduction}
Ever since the AdS/CFT correspondence was proposed in \cite{Maldacena:1997re}, the supersymmetric AdS solutions of supergravity theory have been extensively studied as the gravity duals to the superconformal field theories formulated on the boundaries of AdS space.
However, for the case of AdS$_6$, which would be the gravity dual of five-dimensional SCFT, there is a relatively small number of solutions known in the literature. The first one was found in massive type IIA supergravity by Brandhuber and Oz \cite{Brandhuber:1999np}, which is a warped product of AdS$_6$ space and a half of $S^4$. Two other solutions were obtained by abelian/non-abelian T-dual transformations of the Brandhuber-Oz solution \cite{Cvetic:2000cj, Lozano:2012au}. The systematic searches for AdS$_6$ solutions also have been done. It was proved that the Brandhuber-Oz solution is the unique one in massive IIA theory \cite{Passias:2012vp}. In type IIB supergravity, Apruzzi et al. studied the most general supersymmetric conditions and obtained a set of BPS equations, which are coupled partial differential equations \cite{Apruzzi:2014qva}. They also showed that AdS$_6$ solution dose not exist in M-theory.
In \cite{Kim:2015hya}, we revisited the IIB solutions and studied $D=4$ effective theory obtained by a dimensional reduction on AdS$_6$ space to get more insight on the AdS$_6$ solutions and their properties. We discovered that the unexpected symmetry $SL(3,\mathbb{R})/SO(2,1)$ came out in a dimensional reduction.
In this note we continue to study the hidden symmetry structure of AdS$_6$ solutions in type IIB supergravity theory.

We consider general AdS$_6$ and the metric is written as follows, 
\be
ds_{10}^2=e^{2U}\, L^2\, ds^2\left(\textrm{AdS}_6\right)+e^{-6U}\,ds^2\left(M_4\right).
\ee
Here the warp factor $U$ is a function on $M_4$, and $L$ is a book-keeping parameter and sets the curvature
radius scale of AdS$_6$. We rescale the four-dimensional space by $e^{-6U}$ to obtain an Einstein frame action in the dimensionally reduced theory.
Among the fields of IIB theory
the five-form flux $F_5$ is set to be zero and the complexified three-form flux $G$ will be 
dualized to two real scalars $f$ and $g$ in $M_4$. In addition, there is axio-dilaton field $C+ie^{-\phi}$. After all there are five scalars on $M_4$.

Since all the fields should be independent of coordinates of AdS$_6$, we can dimensionally reduce
the $D=10$ equations down to $D=4$ on $M_4$. The equations are summarized by a non-linear sigma model of five scalar fields interacting with gravity. In \cite{Kim:2015hya}, we pointed out that the target
space is in fact coset $SL(3,\mathbb{R})/SO(2,1)$. More concretely, the Lagrangian is 
\begin{align}\label{action-original}
{\cal L}=&\sqrt{g_4}  \Big[ R
-24 (\partial U)^2
-\frac{1}{2}(\partial \phi)^2-\frac{1}{2}e^{2\phi}(\partial C)^2\nn \\
	&\qquad +\frac{1}{2}e^{-12U-\phi}(\partial f)^2+\frac{1}{8}e^{-12U+\phi}(\partial g+ 2C\partial f)^2 -\frac{30}{L^2}e^{-8U}
\Big], \\\label{action-symm}
=& \sqrt{g_4} \left[ R-P_{\mu i j}P^{\mu i j}-\frac{30}{L^2} e^{-8U}\right].
\end{align} 
Our notations and conventions for the coset construction can be found in the appendix \ref{coset}. Note
also the scalar potential, which originates from the non-vanishing curvature of the AdS$_6$ space. This scalar potential depends on only one scalar field $U$ and breaks the global symmetry  $SL(3,\mathbb{R})$ into a certain sub-algebra which is so-called $A_{5,40}\cong sl(2,\mathbb{R})\ltimes \mathbb{R}^2$ according to the classification in \cite{Patera:1976ud}.  
 
The scalar $U$ is what we usually call ``breathing-mode" in Kaluza-Klein reduction analysis. In general, the breathing mode resides in a massive multiplet and does not appear in the supersymmetric truncation. Our approach here is to cover all AdS$_6$ solutions in a consistent way, so we keep $U$ 
since it is a singlet of AdS$_6$ isometry group.

We should emphasize that our four-dimensional theory above is {\it not} a supersymmetric theory 
on its own. Of course as it stands it is purely bosonic, but even if we include all fermionic fields which 
are singlets with respect to AdS$_6$, the action cannot be made supersymmetric. One way of 
seeing it is that the number of scalar fields is odd, so they cannot parametrize a K\"ahler manifold.
However the action is equipped with the Killing spinor equations which are inherited from the ten-dimensional supergravity transformation rules. Any solution of the equations from action \eqref{action-symm} gives rise to an exact $D=10$ solution, supersymmetric or otherwise. 
And as usual, if the solution admits a non-trivial solution to the associated Killing spinor equations,
the uplifted $D=10$ solution is supersymmetric. 
 
Superficially the action \eqref{action-symm} is similar to the bosonic sector of gauged supergravity which
can be obtained through Kaluza-Klein reduction on curved internal spaces. The scalar fields parametrize
a coset, and the curvature $L^{-2}$ plays a role of coupling constant for the scalar potential. But
it is rather puzzling that there are no gauge fields here in \eqref{action-symm}. 
Usually in extended supergravity models
there are a number of vector fields, which through the gauging process become non-abelian gauge 
fields and at the same time scalar potential terms are introduced to the action. Gauging process 
induces new terms in the supersymmetry transformation rule, and also changes the integrability conditions
which should be consistent with the Einstein and matter field equations. We recall that, in gauged 
supergravity, a covariant derivative acting on the dilatino variation rule leads to gauge field equations
as well as the scalar field equations.\footnote{A classic example of such a computation can be found e.g. in the appendix B of \cite{Pernici:1984zw}, for $D=7$ maximal supergravity.} And in the gauge field equations, the derivative of a certain scalar potential term plays a role of composite current which source the gauge field. Naively, we expect to have
a problem since with \eqref{action-symm} we have scalar potential but no vector field. This line of 
thought motivated us to check the integrability conditions of Killing spinor equations for \eqref{action-symm} carefully. In that process, as a by-product we confirm that \eqref{action-symm} can be in fact
generalized to arbitrarily larger cosets with general signature, $SL(n,\mathbb{R})/SO(p,q)$ with $p+q=n$.

The main purpose of this paper is thus to clarify carefully the hidden symmetry structure and requirement of AdS$_6$ solutions of type IIB supergravity. We focus on the scalar potential of \eqref{action-symm}, and first rewrite the associated Killing spinor equations in a $SL(3,\mathbb{R})/SO(2,1)$-covariant way.
By examining the integrability conditions of the Killing spinor equations, we identify 
the covariant form of
the scalar potential and the unbroken symmetry of the theory.
This study might shed new light on the understanding of the geometric structure of AdS$_6$ solutions in IIB supergravity.

Based on this, we generalize this analysis to more general cosets and classify the four-dimensional theories admitting 
Killing spinor equations.
It enables us to investigate a class of non-supersymmetric $D=4$ gravity theory, coupled to a non-linear sigma model with a non-trivial scalar potential, using BPS equations.

We expect that the similar constructions might be possible in a similar setup, for example, for gravity solutions containing $d+1$ dimensional AdS or Minkowski space. Then, the coset $G/H$ may emerge as a symmetry of the solution space in dimensional reduction of the theory. From the view point of a lower dimensional theory, we would like to emphasize again that it is not included in gauged supergravity theory due to the presence of the breathing mode and the absence of the gauge fields. Even though the theory is purely bosonic, it is interesting that there is a set of associated Killing spinor equations which enjoy the coset symmetry and are compatible with the field equations. Hence, it is desirable to study the various examples to have a concrete understanding and this paper may serve as a starting point for further investigations.

This paper is organized as follows. In section \ref{nlsm}, we study the hidden symmetry structure of IIB AdS$_6$ solutions. Section \ref{general} discusses the classification by the generalized Killing spinor equations. The technical details are relegated to the appendices. In appendix \ref{AdS6M}, we present a similar construction and analysis for AdS$_6$ solutions in $D=11$ supergravity.
\section{Coset non-linear sigma model}\label{nlsm}
\subsection{Scalar potential}\label{S-pot}
We start by illustrating the symmetry structure of the scalar potential of our theory, which was not manifest in the form of Lagrangian written as \eqref{action-symm}. The basic building block of the non-linear sigma model is the coset representative $\cal{V}$. Hence, our goal is to express the scalar potential in terms of ${\cal V}$. We first rewrite the Killing spinor equations of our system \eqref{KSE-scalar} into a covariant form, which respect $SL(3,\mathbb{R})/SO(2,1)$. Then we study the integrability conditions of these Killing spinor equations. They yield the equations of motion and enable us to identify the scalar potential in terms of the coset representative.

With appropriate linear combinations and rescalings, which are explained in detail in the appendix \ref{KSE}, the Killing spinor equations can be re-phrased in a more covariant way. The first set of Killing spinor equations, which we call ``gravitino variation", can be written as
\begin{align}
 \left( \begin{array}{c} \delta \psi_{\mu +} \\ \delta \psi_{\mu -}
         \end{array}\right)
 \equiv  \left( \begin{array}{cc}
            D_\mu             & S \gamma_\mu \\
            S \gamma_\mu &  \bar{D}_\mu      
     \end{array} \right)   
     \left( \begin{array}{c} \xi_+ \\ \xi_-
            \end{array}   \right),      \label{KS1}
\end{align}
and the second set, which we call ``dilatino variation", can be written as
\begin{align}
 \left( \begin{array}{c} \delta \lambda_{i +} \\ \delta \lambda_{i -}
         \end{array}\right)
 \equiv  \left( \begin{array}{cc}
            M_{ij}\bar{\Gamma}^j            &  P_{\mu ij}\gamma^{\mu}\bar{\Gamma}^j   \\
             P_{\mu ij}\gamma^{\mu}\Gamma^j &  M_{ij}\Gamma^j       
     \end{array} \right)   
     \left( \begin{array}{c} \xi_+ \\ \xi_-
            \end{array}   \right),    \label{KS2}
\end{align}
where
\begin{align}
D_\mu &= \nabla_\mu +\frac{1}{4} Q_{\mu ij}\Gamma^{ij}, \qquad
\bar{D}_\mu = \nabla_\mu +\frac{1}{4} Q_{\mu ij}\bar{\Gamma}^{ij}, \nn \\
\Gamma^i &=(\tau^2,\tau^1,-i\tau^3),
          \qquad \quad\bar{\Gamma}^i=(\tau^2,\tau^1,i\tau^3), 
          \qquad \xi_\pm = \left( \begin{array}{c} \xi_{1 \pm} \\  \xi_{2 \pm}
         \end{array}\right). 
\end{align}
Here $\nabla_\mu$ is an ordinary covariant derivative and
$\tau^i$ are Pauli matrices. We have two different representations of gamma matrices $\Gamma^i$ and $\bar{\Gamma}^i$.
A scalar $S$ and a matrix $M_{ij}$ are 
\begin{equation}\label{AdS-SM}
S=\dfrac{3i}{2L} e^{-4U}, \qquad M_{ij}=\frac{2 i}{L} e^{-4U}\textrm{diag}(1,1,2).          
\end{equation}
 Their presence is of course due to the non-trivial scalar potential in our theory,
and we may regard them as an analogue of the $T$-tensor or its combinations in ordinary gauged supergravity theories.

Now let us turn to the integrability conditions of the Killing spinor equations \eqref{KS1} and \eqref{KS2}. We encounter two types of the integrability conditions. The first one is the gravitino-gravitino integrability condition \eqref{gg-integ} and the second one is the dilatino-gravitino condition \eqref{dg-integ}. To uncover the symmetry structure of the scalar potential, we examine the integrability conditions with general, arbitrary $S$ and $M_{ij}$. The detailed calculation is given in the appendix \ref{IC}. These integrability conditions give rise to the Einstein equation and the scalar equation of motion
as
\begin{equation}
\begin{gathered}
R_{\mu\nu}-\dfrac{1}{2} R\, g_{\mu\nu}
    -\left( P_{\mu ij} P_{\nu}^{\phantom{\nu} ij}-\dfrac{1}{2} g_{\mu\nu} P_{\rho ij} P^{\rho ij}\right) 
    +g_{\mu\nu} \left(-12S^2 +\dfrac{1}{2}M_{ij}M^{ij}\right)=0, \\
D_\mu P^\mu_{\phantom{\mu}ij} -4S M_{ij}-2S \left( KMK \right)_{ij}
             +\dfrac{1}{2} M_{ij}\textrm{Tr} \left( MK \eta\right)=0.
\end{gathered}\label{eom}
\end{equation}
From the Einstein equation, one can easily read off the scalar potential as
\be\label{pot}
V=-24S^2 +M_{ij}M^{ij}.
\ee
Here $S$ and $M_{ij}$ are solutions to the following equations
\begin{align}\label{cancellation}
&\partial_\mu S + \dfrac{1}{4}M_{ij}P_\mu^{\phantom{\mu} ij}=0,\nn\\
&\partial_\mu M_{ij}+ \left(Q_\mu \eta M -M\eta Q_\mu \right)_{ij}
        +\dfrac{1}{2} \textrm{Tr} \left(P_\mu K \eta\right)M_{ij}
        +2S P_{\mu ij}-2S \left(K P_\mu K \right)_{ij}=0,\nn\\
&SQ_\mu-KSQ_\mu K+\dfrac{1}{4} \left( P_\mu\eta M-M\eta P_\mu\right)
-\dfrac{1}{4} K\left( P_\mu\eta M-M\eta P_\mu\right)K=0, \nn\\
&P_\mu \eta M-M\eta P_\mu +K\left( P_\mu \eta M-M\eta P_\mu \right)K=0,\nn\\
&MK-KM=0.
\end{align}
These five equations appear in the integrability conditions \eqref{gg-integ} and \eqref{dg-integ}
and are required for compatibility with the field equations.
We have found a solution to \eqref{cancellation}
\begin{align}\label{solSM}
S&= \dfrac{3}{4}\,\dfrac{\alpha}{{\cal{V}}_{33}}, \qquad
M_{ij}= \dfrac{\alpha}{{\cal{V}}_{33}}\, \textrm{diag}(1,1,2). 
\end{align}
Here $\alpha$ is an integration constant and ${\cal{V}}_{ii}$ is an ($i,i$)-component of the coset representative $\cal{V}$. The scalar potential $V$ at hand turns out to be related to $(3, 3)$-component of $\cal{V}$ through \eqref{pot} and \eqref{solSM}.

Now we rephrase the main result of this subsection. Let us suppose that we are given the Killing spinor equations \eqref{KS1} and \eqref{KS2}, which can be written in terms of the $SL(3,\mathbb{R})/SO(2,1)$ coset representative $\cal{V}$. A scalar $S$ and a matrix $M_{ij}$ are given by \eqref{solSM}, which implies that the global $SL(3,\mathbb{R})$ symmetry is broken to $sl(2,\mathbb{R})\ltimes \mathbb{R}^2$ due to a non-trivial scalar potential as we will show in the next subsection.
Then, we can construct a four-dimensional theory admitted by the Killing spinor equations as
\be\label{action-gen}
{\cal{L}}=\sqrt{g_4} \Big( R -P_{\mu ij} P^{\mu ij}-\left( M_{ij}M^{ij}-24S^2\right) \Big).
\ee 
If we substitute the specific solution \eqref{AdS-SM}, the Lagrangian \eqref{action-gen} reduce to our starting point Lagrangian \eqref{action-symm}, whose equations of motion is \eqref{eom}.
\subsection{Group action}\label{groupaction}
We have shown that the scalar potential depends only on the $(3, 3)$-component of the coset representative. As a consequence, the global $SL(3,\mathbb{R})$ symmetry of the target space is broken to a non-trivial sub-algebra, under which ${\cal{V}}_{33}$ is invariant.
In this section, we directly apply the group transformations on the coset representative and elucidate unbroken symmetry which is preserved by the whole theory.

The coset representative ${\cal{V}}$ transforms as
\be
 {\cal{V}} \longrightarrow  K\, {\cal{V}}\, G,
\ee
under global G and local K transformations. Infinitesimally, it transforms as
 \be
 {\cal{V}} \longrightarrow   {\cal{V}} + \alpha\left({\Phi^I}\right)\, k\, {\cal{V}} + {\cal{V}}\, g
\ee
where $g$'s and $k$'s are generators of G and K, respectively. $\alpha$'s are compensators and functions of $\Phi^I$, which are the coordinates of coset space G/K.

Now we study group actions on an $SL(3,\mathbb{R})/SO(2,1)$ coset representative ${\cal{V}}$ in the Borel gauge \eqref{corep}.
Under the action of Cartan generators $h_1,\, h_2$ and positive root generators $e_1,\, e_2,\, e_3$ of $SL(3,\mathbb{R})$, the coset representative remains in the Borel gauge. Hence there is no need for compensators. On the other hand, when we transform ${\cal{V}}$ with negative root generators $f_1,\, f_2,\, f_3$, we need the non-trivial compensators to restore the Borel gauge. 
The transformations with appropriate compensators are written explicitly in the table below
\begin{center}
\begin{tabular}{r|r}
 &\multicolumn{1}{c}{ $\delta {\cal{V}}$} \\
 \hline
$f_1$ & $  e^{-\phi}\,  k_1\, {\cal{V}} + \, {\cal{V}}\,    f_1$ \\
$f_2$ & $- e^{6U+\phi/2}\,  k_2\, {\cal{V}} + \,  {\cal{V}}\,   f_2 $\\
$f_3$ & $\left(e^{-\phi}\,  f\,  k_1+  e^{6U+\phi/2}\,  C\,  k_2-e^{6U-\phi/2}\, k_3 \right) {\cal{V}}    + \,  {\cal{V}}\,   f_3 $
\end{tabular}
\end{center}
where $k_1,\, k_2,\, k_3$ are $SO(2,1)$ generators.
The scalar kinetic terms are invariant under the action of all the generators of $SL(3,\mathbb{R})$. However,
the scalar potential \eqref{pot}, i.e. ${\cal{V}}_{33}$ is invariant under only some of the generators.
Explicit computations show that our theory is invariant under the action of generators $h_1,\, e_1,\, e_2,\, e_3$ and $f_1$ only.

Now let us focus on these five generators. Under the action of  $e_1,\, h_1,\, f_1,\, e_2,\, e_3$, each scalar field transforms as
\begin{center}
\begin{tabular}{l|cccc}
 & $\delta e^\phi$ & $\delta C$ & $\delta f$ & $\delta g$\\
\hline
$e_1$ & & $1$
         & & $ -2\,f$ \\ 
$h_1$ & $\sqrt{2}\,  e^\phi$ 
         & $- \sqrt{2}\,  C$ 
         & $ 1/\sqrt{2}\,f$ 
         & $-1/\sqrt{2}\,g$ \\
         $f_1$ & $ 2\,  C\, e^\phi$   
         & $ e^{-2\phi}-C^2 $
         & $ -1/2 \,g$ \\
$e_2$ & & &$ 1$ \\
$e_3$ & & & & $  2$ \\ 
\hline
\end{tabular}
\end{center}
The first three transformations under the action of the generators $e_1,\, h_1$ and $f_1$ exactly correspond to $SL(2,\mathbb{R})$ transformations of type IIB supergravity, which transform\footnote{The definition of 2-from potential $C_2$ and $B_2$ can be found in \eqref{com3form}.}
\begin{align}
\tau \rightarrow \dfrac{p \tau + q}{r \tau +s}, \qquad
\left(
\begin{array}{c}
C_{2}\\
B_{2}
\end{array}
\right) \rightarrow
\left(
\begin{array}{cc}
p & q\\
r & s
\end{array}
\right) 
\left(
\begin{array}{c}
C_{2}\\
B_{2}
\end{array}
\right),
\end{align}
where $\tau= C + i\, e^{-\phi}$ and $p\,s-q\,r=1$,
with the transformation matrices
\begin{align}
\left(
\begin{array}{cc}
 1 & a \\
 0 & 1
\end{array}
\right), \quad
\left(
\begin{array}{cc}
 1-\dfrac{a}{\sqrt{2}} & 0 \\
 0 &1+\dfrac{a}{\sqrt{2}}
\end{array}
\right), \quad
\left(
\begin{array}{cc}
 1 & 0 \\
 a & 1
\end{array}
\right),
\end{align}
respectively. Here $a$ is a transformation parameter. The last two transformations are constant shift in $f$ and $g$.

This result is consistent to the previous analysis.
In \cite{Kim:2015hya}, we showed that the scalar potential is invariant under the action of five Killing vectors 
$K^1,\, K^3,\, K^4,\, K^6$ and $K^8$, which generate a non-trivial algebra $A_{5,40}\cong sl(2,\mathbb{R})\ltimes \mathbb{R}^2$.
Using the following relations
\begin{align}
\delta_M \cal{V}&=  \alpha_M^{\phantom{M}i}\left({\Phi^I}\right)\, k_i\, {\cal{V}} + \, {\cal{V}}\, g_M, \nn\\
&=K_M {\cal{V}}
\end{align}
where $K_M,\,M=1, \cdots, 8$ are the Killing vectors (E.1) in \cite{Kim:2015hya},
we can identify the eight Killing vectors with generators of SL($3,\mathbb{R}$) as
\begin{center}
\begin{tabular}{cccccccc}
$K^1$ & $K^2$ & $K^3$ &$K^4$ & $K^5$ & $K^6$ & $K^7$ &$K^8$\\
\hline
$h_1$ & $-h_2$ & $-\frac{1}{2}e_1$ & $-2f_1$ & $2f_3$ & $\frac{1}{2}e_3$ & $f_2$ &$e_2$ 
\end{tabular}
\end{center}
More precisely, the five Killing vectors $(K^1, K^3, K^4, K^6, K^8)$ correspond to the generators $(h_1, -\frac{1}{2}e_1, -2f_1, \frac{1}{2}e_3, e_2)$.
Therefore, it is the $SL(2,\mathbb{R})$ symmetry of type IIB supergravity and the trivial gauge transformations of the dualized scalars that account for the appearance of the non-trivial algebra $A_{5,40}$ in \cite{Kim:2015hya}.
\subsection{Examples}\label{ex}
We have shown how to make the symmetry $SL(3,\mathbb{R})/SO(2,1)$ manifest in AdS$_6$ solutions of type IIB supergravity. In this section, we identify the $SL(3,\mathbb{R})/SO(2,1)$ symmetry  of the known solutions. 

The first AdS$_6$ solution in type IIB supergravity was obtained by taking abelian T-dual transformation of the Brandhuber and Oz solution \cite{Cvetic:2000cj, Lozano:2013oma}. The solution is given by (see eq. (A.1) in \cite{Lozano:2013oma})
\begin{align}\label{AT-sol}
ds^2 &= \frac{1}{4}\,W^2\, L^2 \left( 9\,ds^2(AdS_6)+4\, d\theta^2 +\sin^2\theta \left(d\phi_1^2+\sin^2\phi_1\, d\phi_2^2\right)+\frac{16}{W^4\, L^6\, \sin^2\theta}\,d\phi_3^2\right),\nn \\
B&= -\cos\phi_1\, d\phi_2 \wedge d\phi_3, \nn\\
F_3&= \frac{5}{8}\, L^4\, (m \cos\theta)^{1/3}\, \sin^3\theta\, \sin\phi_1\, d\theta \wedge d\phi_1 \wedge d\phi_2, \nn\\
F_1&= m\, d\phi_3, \nn\\
e^\phi&= \frac{4}{3\,L^2\, (m \cos\theta)^{2/3}\, \sin\theta},
\end{align}
where $W=(m \cos\theta)^{-1/6}$ and $m$ is the Romans' mass in the original massive IIA theory.
We rewrite the metric in the Einstein frame
\be
ds_E^2 = \frac{\sqrt{3}}{8}\,L^3\,( \sin\theta)^{1/2}\left( 9\,ds^2(AdS_6)+4\, d\theta^2 +\sin^2\theta \left(d\phi_1^2+\sin^2\phi_1\, d\phi_2^2\right)+\frac{16}{W^4\, L^6\, \sin^2\theta}\,d\phi_3^2\right),
\ee
and identify the warp factor as
\be
e^{2U}= \frac{9\sqrt{3}}{8}\,L\,( \sin\theta)^{1/2}.
\ee
The axion $C$ and the dilaton $e^\phi$ are given by solution where $F_1=dC$. We can easily obtain the dualized scalars $f$ and $g$ by using \eqref{com3form}.
By substituting solutions, i.e. these five scalar fields $U,\, \phi,\, C,\, f$ and $g$ into the coset representative \eqref{corep}, 
we obtain
\begin{align}
 {\cal{V}}=\left( \begin{array}{ccc}
  \dfrac{16\, \csc\theta}{27\,L^2\, (m \cos\theta)^{1/3}} 
  &  \quad \dfrac{16\, m\,\phi_3\, \csc\theta}{27\,L^2\, (m \cos\theta)^{1/3} }  
  & \quad \dfrac{27}{128\,L^2}\left(27\, L^4\, \cot\theta-\dfrac{80\, m\, \phi_3^{\phantom{3}2}\,\csc\theta}{(m \cos\theta)^{1/3} }\right)\\
  0 & \quad \dfrac{4}{9}\,(m\, \cos\theta)^{1/3} 
  & \quad -\dfrac{405}{16}\,\phi_3\, (m \cos\theta)^{1/3} \\
  0 & \quad 0 & \quad \dfrac{243}{64}\,L^2\,\sin\theta
  \end{array}\right).
\end{align}
This coset representative shows the $SL(3,\mathbb{R})/SO(2,1)$ symmetry manifestly and have all the informations to reconstruct the abelian T-dual AdS$_6$ solution \eqref{AT-sol}.
Here we record the explicit expression for $P_\mu$,
\begin{align}\label{Tsol-P}
P_\theta =\left(
                 \begin{array}{ccc}
                  \cot\theta-\dfrac{1}{3}\tan\theta & 0 & -\dfrac{1}{\sin\theta} \\
                  0 & \dfrac{1}{3} \tan\theta &0 \\
                  -\dfrac{1}{\sin\theta} & 0 &\cot\theta 
                 \end{array}
               \right), \qquad
P_{\phi_3} =-\dfrac{2\,m\,(m\, \cos\theta)^{-2/3}}{3\,L^2\, \sin\theta}
             \left(
                 \begin{array}{ccc}
                  0& 1 & 0 \\
                  1 & 0 & 5\cos\theta\\
                  0 & 5\cos\theta & 0
                 \end{array}
               \right).
\end{align}

The second AdS$_6$ solution in type IIB supergravity is a non-abelian T-dual solution obtained by \cite{Lozano:2012au}. Similarly, we can easily read off the five scalar fields and identify the coset representative. However, the explicit expression is not so illuminating. We do not present it here.
\section{Generalization to $SL(n,\mathbb{R})/SO(p,q)$}\label{general}
In section \ref{S-pot}, we have shown that the Killing spinor equations \eqref{KS1} and \eqref{KS2} consist of two parts. The diagonal components of the gravitino variation equation and the off-diagonal components of the dilatino variation equation enjoy the coset $SL(3,\mathbb{R})/SO(2,1)$ symmetry. On the other hand, the remaing part of the Killing spinor equations show that the symmetry is broken to $sl(2,\mathbb{R})\ltimes \mathbb{R}^2$. Based on this symmetry structure, we have constructed a four-dimensional theory \eqref{action-gen}. 


In this section, we formally generalize this construction to respect the coset symmetry $SL(n,\mathbb{R})/SO(p,q)$ with arbitrary $n$ and signature. Since our construction is covariant to the coset symmetry, the form of the main equations is basically same. First, the construction of the $SL(n,\mathbb{R})/SO(p,q)$ coset representative $\cal{V}$, which is now an $n \times n$ matrix, is very straightforward.
We also define the invariant metric of $SO(p,q)$ as
\be
\eta =\textrm{diag}(\underbrace{1,\cdots,1}_p,\underbrace{-1\cdots,-1}_q),
\ee
where $p+q=n$, and introduce a matrix $K$ 
\be
K^i_{\phantom{i}j}= \textrm{diag}(\underbrace{1,\cdots,1}_{n-1},-1),
\ee
such that  $\Gamma^i =K^i_{\phantom{i}j}\,\bar{\Gamma}^j$ where $i=1, \cdots, n$.
Then, we have the generalized version of the Killing spinor equations \eqref{KS1} and \eqref{KS2}. Examining the integrability conditions give rise to the generalized equations of motion \eqref{eom}. Finally, we find solutions to eq. \eqref{cancellation}, which are now written in terms of $n \times n$ matrices, as
\begin{align}\label{solSM-gen}
S= \dfrac{n}{4}\,\dfrac{\alpha}{{\cal{V}}_{nn}}, \qquad
M_{ij}=\dfrac{\alpha}{{\cal{V}}_{nn}} \widetilde{M}_i^{\phantom{i}k}\,\eta_{kj}.
\end{align}
where 
\be
\widetilde{M}_i^{\phantom{i}j}= \textrm{diag}(\underbrace{1,\cdots,1}_{n-1},-(n-1)).
\ee
As a result, we can obtain new four-dimensional theories by plugging the solutions \eqref{solSM-gen} into the Lagrangian \eqref{action-gen}.
This generalization implies that we can classify and construct the theories admitting the generalized Killing spinor equations \eqref{KS1} and \eqref{KS2}. More precisely, we can construct four-dimensional gravity theory coupled to a non-linear sigma model of $(n^2+n-2)/2$  scalar fields with a certain class of scalar potential. The non-linear sigma model parametrize the coset space $SL(n,\mathbb{R})/SO(p,q)$ and the scalar potential is determined by $(n, n)$-component of the coset representative $\cal{V}$. The $SL(n,\mathbb{R})$ global symmetry is broken to a sub-algebra $sl(n-1,\mathbb{R})\ltimes \mathbb{R}^{n-1}$ which preserve this component.

Let us consider the simplest example, which is $SL(2,\mathbb{R})/SO(1,1)$. The Lagrangian can be written 
as follows,
\begin{align}
{\cal L}=&\sqrt{g_4}  \Big[ R
-\frac{1}{2}(\partial A)^2+\frac{1}{2}e^{-2A}(\partial B)^2 +4\, \alpha^2\, e^{-A} \Big]. \label{action-ex}
\end{align}	
At first sight, this Lagrangian looks very different from \eqref{action-original}. It has only two scalar fields. Furthermore, this theory does not have any higher dimensional origin, in contrast to
\eqref{action-original}. However, these two theories can be classified into one category. They share the key property that they can be constructed by the Killing spinor equations of the same type .

 To summarize, if we are given a coset representative $\cal{V}$, which respect $SL(n,\mathbb{R})/SO(p,q)$ symmetry, we can compute $P_{\mu ij},\, Q_{\mu ij},\, S$ and $M_{ij}$ by \eqref{PQ} and \eqref{solSM-gen}. Then, we can easily write down the generalized Killing spinor equations  \eqref{KS1}, \eqref{KS2} and construct the compatible four-dimensional theories \eqref{action-gen}. The important feature of our theories is that they have Killing spinor equations although the theory is not a supergravity theory itself.\footnote{The non-supersymmetric theories admitted by the Killing spinor equations were referred to as ``fake supergravity" \cite{Freedman:2003ax} and ``pseudo-supersymmetry"\cite{Lu:2011zx}.}
 With these Killing spinor equations, one may study a BPS sector of theory and search for the bosonic BPS equations. 
\section{Discussions}
In this note we have studied the appearance of $SL(3,\mathbb{R})/SO(2,1)$ in AdS$_6$ solutions of type IIB supergravity \cite{Kim:2015hya}.
We managed to rewrite the associated Killing spinor equations also in a form compatible with the symmetry
of the coset. Studying their integrability conditions enabled us to understand the scalar potential in terms of the $(3, 3)$-component of the coset representative ${\cal{V}}$. As a consequence, the theory is invariant under the action of the some generators, which leave the $(3,3)$-component of the coset representative fixed. We have shown that they satisfy a non-trivial algebra $sl(2,\mathbb{R})\ltimes \mathbb{R}^2$, which exactly corresponds to the $SL(2,\mathbb{R})$ symmetry of type IIB supergravity and trivial gauge transformations of the dualized scalar fields, respectively.

Based on these analyses, we have rewritten the two known IIB AdS$_6$ solutions according to the symmetry structure. 
Given the solutions which respect the coset symmetry, it is well known that a new solution may be generated by the action of the group transformations on the coset representative of a known solution (See, for example, \cite{Bouchareb:2007ax, Compere:2009zh}). 
Original motivation of this work was in fact to find new solutions, employing such ``solution generating technique". But unfortunately we were not able to do it, since the scalar potential breaks $SL(3,\mathbb{R})$ to $SL(2,\mathbb{R})$, which is just the familiar S-duality of the type IIB supergravity theory. 

We also have constructed a class of four-dimensional gravity theories coupled to scalar fields. Their kinetic terms are described by the coset $SL(n,\mathbb{R})/SO(p,n-p)$ non-linear sigma model and scalar potential is determined by $(n, n)$-component of the coset representative. These non-supersymmetric theories have nice properties that they are endowed with the Killing spinor equations.

 As opposed to ordinary gauged supergravity theories, our model does not have any gauge fields to promote the global symmetry to local ones. However, by introducing the analogues of $T$-tensor in ordinary gauged supergravity, the global symmetry is broken to its sub-algebra and the scalar potential can be generated. 
More specifically, we have introduced a scalar $S$ and a matrix $M_{ij}$, which can be determined in terms of the coset representative as \eqref{solSM-gen}, such that the scalar potential becomes \eqref{pot}. 
In the appendix \ref{AdS6M}, a similar analysis was done via a dimensional reduction of D=11 supergravity on AdS$_6$. In that case, two vectors $S_i$ and $T_i$ are introduced instead. With a specific solution \eqref{solST-gen}, the scalar potential becomes \eqref{pot-mth}. 
In each case, the analogues of $T$-tensor and the explicit expressions of the scalar potentials are different. However, their dependences on the coset representative and the unbroken symmetries are same. In other words, the scalar potentials can be completely determined by using only one-component of the diagonal elements of the coset representatives. Hence, the global symmetry $SL(n,\mathbb{R})$ is broken to $sl(n-1,\mathbb{R})\ltimes \mathbb{R}^{n-1}$ algebra in both cases. 
One may ask whether this choice of a sub-algebra is unique or not. At present, we do not fully understand the mechanism of the global symmetry breaking beyond the examples which we have presented in this paper. For this reason, it will be interesting to examine all the possible subgroups which generate consistent theories with the non-trivial scalar potential. We hope to study this process in other examples and find a general, systematic method in a near future.
\acknowledgments
We would like to thank Minwoo Suh for collaboration on a related work, and also for discussions and
comments on the manuscript. This work was partly done while we were both attending
the APCTP focus program ``Duality and Novel Geometry in M-theory". We are grateful to APCTP for hospitality. This research is supported by NRF grants 2015R1D1A1A09059301 (NK), 2016K2A9A2A08003745 (NK) and 2013R1A1A2064824 (HK).
\appendix

\section{$SL(3,\mathbb{R})/SO(2,1)$ coset representative}\label{coset}
The coset representative in the Borel gauge is obtained by exponentiating the Cartan generators and the positive root generators. The explicit form of $SL(3,\mathbb{R})/SO(2,1)$ coset representative in terms of the five scalar fields is
\begin{align}\label{corep}
 {\cal{V}}
 &=e^{\frac{1}{\sqrt{2}}\,\phi\, h_1} e^{-2\sqrt{6}\,U\, h_2} e^{C\, e_1} e^{f\, e_2} 
    e^{\frac{1}{2}\, g\,e_3}, \nn\\
 &=\left( \begin{array}{ccc}
  e^{-2U+\phi/2} &\quad e^{-2U+\phi/2} C &  \quad \frac{1}{2}e^{-2U+\phi/2} \left(g+2 C f\right) \\
  0 & \quad e^{-2U-\phi/2} & \quad e^{-2U-\phi/2} f \\
  0 & \quad 0 & \quad e^{4U}
  \end{array}\right).
\end{align}
Here $h_1, h_2$ are Cartan generators and $e_i$'s are positive root generators of $SL(3,\mathbb{R})$. In addition, three negative root generators are $f_i =e^T_i$. The explicit form of the generators are given by eq. (E.2) in \cite{Kim:2015hya}. The generators of $SO(2,1)$  are defined by
\be
k_1 =e_1 -f_1, \quad k_2 =e_2 +f_2, \quad k_3 =e_3 +f_3.
\ee
We introduce a Lie algebra-valued one-form as
\be\label{PQ}
{\cal{V}}_{i}^{\phantom{(i}m} \partial_\mu({\cal{V}}^{-1})_m^{\phantom{m}k}\, \eta_{kj}=P_{\mu(ij)}+Q_{\mu[ij]}. 
\ee
Here $P_{\mu ij}$ is an orthogonal complement of $so(2,1)$ in $sl(3,\mathbb{R})$ and $Q_{\mu ij}$ behaves as a composite $SO(2,1)$ gauge field
\be
D_{\mu}P_{\nu ij} =\nabla_\mu P_{\nu ij}+[Q_\mu , P_\nu]_{ij}.
\ee
The definition of $P_\mu$ and $Q_\mu$ \eqref{PQ} give the following integrability relations
\begin{gather}
D_{\,[\mu}P_{\nu ]}=0,\nn\\
\nabla_{\,[\mu}Q_{\nu]} + P_{\,[\mu} P_{\nu]}+ Q_{\,[\mu} Q_{\nu]}=0.
\end{gather}
The two real real scalars $g$ and $f$ can be obtained by dualizing the complex three-form flux $G$.\footnote{Hodge dual is taken with respect to the rescaled four-dimensional space.} Here we also write down the conventions for real three-form fluxes and two-form potentials.
Their relations are useful when we study the $SL(2,\mathbb{R})$ symmetry of type IIB supergravity in \ref{groupaction} and the abelian T-dual AdS$_6$ solution in \ref{ex}.
\begin{align}\label{com3form}
* \textrm{Re}G&=\dfrac{1}{2}e^{-12U+\phi/2} (dg+2C\,df)=e^{-\phi/2}*H_3
=-e^{-\phi/2}* dB_{2},\\
* \textrm{Im}G&= e^{-12U-\phi/2} df=-e^{\phi/2}*F_3
=e^{\phi/2} \left(C *dB_{2}-*dC_{2} \right).\nn
\end{align}

\section{Killing spinor equations}\label{KSE}
In this section, we reorganize the four-dimensional Killing spinor equations (2.2)-(2.7) in \cite{Kim:2015hya} into a covariant form
with respect to $SL(3,\mathbb{R})/SO(2,1)$ symmetry.
By rescaling the four dimensional metric and spinor $\xi$ as
\be
g_{mn} \longrightarrow e^{-6\,U}g_{mn}, \qquad
\xi \longrightarrow e^{-3/2 \,U}\xi,
\ee
the Killing spinor equations in terms of five scalar fields become\footnote{We call Killing spinor equations  (2.2)-(2.7) in \cite{Kim:2015hya} as $\delta \tilde{\psi}_{\mu\pm},\, \delta \hat{\psi}_{\mu\pm},\,
\delta \tilde{\chi}_\mp, \,\delta \hat{\chi}_\mp, \,
 \delta \tilde{\lambda}_\mp, \, \delta \hat{\lambda}_\mp $.}
\begin{align}\label{KSE-scalar}
-2  \delta \tilde{\lambda}_\pm 
&\equiv \left(-\partial_\mu \phi -i e^\phi \partial_\mu C\right) \gamma^\mu \xi_{2\mp}
  + \left( \frac{1}{4}e^{-6U+\phi/2}(\partial_\mu g +2 C \partial_\mu f)+\frac{i}{2} e^{-6U-\phi/2}\partial_\mu f\right) \gamma^\mu \gamma_5 \xi_{1\mp}, \nn \\
-2   \delta \hat{\lambda}_\pm 
&\equiv \left(-\partial_\mu \phi +i e^\phi \partial_\mu C\right) \gamma^\mu \xi_{1\mp}
  + \left( \frac{1}{4}e^{-6U+\phi/2}(\partial_\mu g +2 C \partial_\mu f)-\frac{i}{2} e^{-6U-\phi/2}\partial_\mu f\right) \gamma^\mu \gamma_5 \xi_{2\mp}, \nn \\
4\,e^{-3U} \delta \tilde{\chi}_\pm
  &\equiv \frac{4 i }{L} e^{-4U}\xi_{1\pm} +4 \partial_\mu U \gamma^\mu \xi_{1\mp}
  + \left( \frac{1}{4}e^{-6U+\phi/2}(\partial_\mu g +2 C \partial_\mu f)+\frac{i}{2} e^{-6U-\phi/2}\partial_\mu f\right) \gamma^\mu \gamma_5 \xi_{2\mp}, \nn \\
4\,e^{-3U} \delta \hat{\chi}_\pm
  &\equiv \frac{4 i}{L} e^{-4U}\xi_{2\pm} +4 \partial_\mu U \gamma^\mu \xi_{2\mp}
  + \left( \frac{1}{4}e^{-6U+\phi/2}(\partial_\mu g +2 C \partial_\mu f)-\frac{i}{2} e^{-6U-\phi/2}\partial_\mu f\right) \gamma^\mu \gamma_5 \xi_{1\mp}, \nn \\
\delta \tilde{\psi}_{\mu\pm}
&\equiv \nabla_\mu \xi_{1\pm}-\frac{3}{2}\partial_\nu U \gamma_\mu \gamma^\nu \xi_{1\pm}+\frac{i}{4} e^\phi \partial_\mu C \xi_{1\pm}\nn \\
&-\frac{3}{8} \left( \frac{1}{4}e^{-6U+\phi/2}(\partial_\nu g +2 C \partial_\nu f)+\frac{i}{2} e^{-6U-\phi/2}\partial_\nu f\right)\gamma_\mu \gamma^\nu \gamma_5 \xi_{2\pm} \nn\\
&+\frac{1}{2} \left( \frac{1}{4}e^{-6U+\phi/2}(\partial_\mu g +2 C \partial_\mu f)+\frac{i}{2} e^{-6U-\phi/2}\partial_\mu f\right) \gamma_5 \xi_{2\pm}, \nn \\
\delta \hat{\psi}_{\mu\pm}
&\equiv \nabla_\mu \xi_{2\pm}-\frac{3}{2}\partial_\nu U \gamma_\mu \gamma^\nu \xi_{2\pm}-\frac{i}{4} e^\phi \partial_\mu C \xi_{2\pm}\nn \\
&-\frac{3}{8} \left( \frac{1}{4}e^{-6U+\phi/2}(\partial_\nu g +2 C \partial_\nu f)-\frac{i}{2} e^{-6U-\phi/2}\partial_\nu f\right)\gamma_\mu \gamma^\nu \gamma_5 \xi_{1\pm} \nn\\
&+\frac{1}{2} \left( \frac{1}{4}e^{-6U+\phi/2}(\partial_\mu g +2 C \partial_\mu f)-\frac{i}{2} e^{-6U-\phi/2}\partial_\mu f\right) \gamma_5 \xi_{1\pm}.
\end{align}
The scalar derivative terms in eq. \eqref{KSE-scalar} can be replaced with the components of $P_{\mu}$ and $Q_{\mu}$. Then, we define $\delta \lambda_{i\pm}, \, \delta \psi_{\mu \pm}$ with the following combinations,
\begin{align}
\delta \lambda_{1\pm}&\equiv \tau^2 \left[
 -\left( \begin{array}{c} 
        \delta \hat{\lambda}_{ \pm} \\ \delta \tilde{\lambda}_{\pm}
 \end{array}\right)
 +2\,e^{-3U}
 \left( \begin{array}{c} 
        \delta \tilde{\chi}_{ \pm} \\ \delta \hat{\chi}_{\pm}
 \end{array}\right) \right], \nn\\
\delta \lambda_{2\pm}&\equiv \tau^1 \left[ \phantom{-}
 \left( \begin{array}{c} 
        \delta \hat{\lambda}_{ \pm} \\ \delta \tilde{\lambda}_{\pm}
 \end{array}\right)
 +2\,e^{-3U}
 \left( \begin{array}{c} 
        \delta \tilde{\chi}_{ \pm} \\ \delta \hat{\chi}_{\pm}
 \end{array}\right) \right],
\end{align}   
and
\begin{align}
\delta \psi_{\mu \pm} \equiv
\left( \begin{array}{c} 
        \delta \tilde{\psi}_{\mu \pm} \\ \delta \hat{\psi}_{\mu\pm}
 \end{array}\right)
 +\frac{3}{2}\gamma_\mu e^{-3U}
 \left( \begin{array}{c} 
        \delta \tilde{\chi}_{ \mp} \\ \delta \hat{\chi}_{\mp}
 \end{array}\right),
\end{align}  
where we devised the form of $\delta \lambda_{i\pm}$ to satisfy the condition 
$\bar{\Gamma}^i\delta \lambda_{i+}=\Gamma^i\delta \lambda_{i-}=0$.
Now we can write down the Killing spinor equations in a covariant form as \eqref{KS1} and \eqref{KS2}.
\section{Integrability conditions}\label{IC}
We study the integrability conditions of Killing spinor equations
\eqref{KS1} and \eqref{KS2} with general $S$ and $M_{ij}$. Let us denote ${\mathcal D}_\mu$ as
\be
{\mathcal D}_\mu \equiv
\left( \begin{array}{cc}
            D_\mu             & S \gamma_\mu \\
            S \gamma_\mu &  \bar{D}_\mu      
     \end{array} \right).
\ee     
 We study the gravitino-gravitino integrability condition and compute
$\gamma_\mu^{\phantom{\mu} \nu\rho} \left[{\mathcal D}_\nu , {\mathcal D}_\rho  \right].$
From the first row, we have
\begin{align}\label{gg-integ}
&  - \dfrac{1}{2}  M^i_{\phantom{i}j}\,\bar{\Gamma}^j\, \gamma_\mu \,  \delta \lambda_{i+}
      +\dfrac{1}{2}  P^{\phantom{\nu} i \phantom{j}}_{\nu \phantom{i} j}\, \gamma^{\nu}\,\Gamma^j\,  \gamma_\mu\, \delta \lambda_{i-}   \nn\\
&+\left(R_{\mu\nu}-\dfrac{1}{2} R g_{\mu\nu}
    -\left( P_{\mu ij} P_{\nu}^{\phantom{\nu} ij}-\dfrac{1}{2} g_{\mu\nu} P_{\rho ij} P^{\rho ij}\right) 
    +g_{\mu\nu} \left(-12S^2 +\dfrac{1}{2}M_{ij}M^{ij}\right) \right) \gamma^{\nu} \xi_+ \nn \\
&-4 \left(\partial_\nu S + \dfrac{1}{4}M_{ij}P_\nu^{\phantom{\nu} ij} \right) \gamma^\nu_{\phantom{\nu} \mu}\xi_{-} \nn \\
%
&-\left(\left(S Q_{\nu ij} +\dfrac{1}{2} \left(P_\nu \eta M\right)_{ij}\right)
-K_i^{\phantom{i}l}\left(S Q_{\nu lk} +\dfrac{1}{2}\left(P_\nu \eta M\right)_{lk}\right)K^k_{\phantom{k}j}\right)
\Gamma^{ij} \gamma^{\nu}_{\phantom{\nu}\mu}\xi_- \nn\\
%
&-\dfrac{1}{2} \left( 
                           \left(P_{\mu} \eta M \right)_{ij}+\left(K P_\mu \eta M K\right)_{ij}
                          \right)\Gamma^{ij} \xi_-,
\end{align}
where we define $K$
\be
K^i_{\phantom{i}j}= \textrm{diag}(1,1,-1),\footnote{$K$ is independent of the choice of the metric signature.}
\ee
such that $\Gamma^i =K^i_{\phantom{i}j}\,\bar{\Gamma}^j$.
The first line vanishes if we impose the Killing spinor equations, i.e. $\delta \lambda _{i \pm}=0$.
The second line yields the Einstein equation. To cancel the third, the fourth and the fifth lines, we have three equations which  $S$ and $M_{ij}$ should satisfy.

The dilatino-gravitino integrability condition is
\begin{align}\label{dg-integ}
\gamma^\mu D_\mu\, \delta\lambda_{i-} 
&=P_{\mu i j }\gamma^\nu \gamma^\mu \Gamma^j \delta \psi_{\nu+}
    + M_{ij} \gamma^\nu \Gamma^j \delta \psi_{\nu-}
    +2S K_i^{\phantom{i}j}\delta\lambda_{j+}
    -\dfrac{1}{2}M_{ij}\Gamma^j \Gamma^k  \delta\lambda_{k+}\\
&+ \left( 
             D_\mu P^\mu_{\phantom{\mu}ij} -4S M_{ij}-2S \left( KMK \right)_{ij}
             +\dfrac{1}{2} M_{ij}\textrm{Tr} \left( MK \eta\right)
       \right) \Gamma^j \xi_+     \nn\\
&+ M_{ij}\left( MK\right)_{kl} \left( \eta^{j[k}\Gamma^{l]}+\dfrac{1}{2}\Gamma^{jkl}\right) \xi_+  \nn\\
&+ \left(
        \partial_\mu M_{ij}+ \left(Q_\mu \eta M -M\eta Q_\mu \right)_{ij}
        +\dfrac{1}{2} \textrm{Tr} \left(P_\mu K \eta\right)M_{ij}
        +2S P_{\mu ij}-2S \left(K P_\mu K \right)_{ij}
      \right) \gamma^\mu \Gamma^j \xi_-,      \nn 
\end{align}
where we used
\be
\left( Q_{\mu}-K Q_\mu K + 2 P_{\mu} K \right)_{[ij]}=0.
\ee
Similarly, the first line vanishes by the Killing spinor equations $\delta \psi_{\mu \pm}=\delta \lambda_{\mu \pm}=0$.
The second line gives the scalar equations of motion. We have two equations for $S$ and $M_{ij}$, which eliminate the third and the fourth lines. 
\section{AdS$_6$ in M-theory}\label{AdS6M}
 In \cite{Apruzzi:2014qva}, the authors showed that supersymmetric AdS$_6 \times M_5$ solutions do not exist in $D=11$ supergravity.
 However, regardless of the existence of the solution, the hidden symmetry structure can be also found in studying AdS$_6$ solutions in M-theory. The analysis is exactly parallel to the IIB AdS$_6$ case. We take a $D=11$ metric as a warped product of AdS$_6$ space and a five-dimensional space as
 \be
 ds_{11}^2 = e^{2U}\, ds^2_{AdS_6}+ e^{-4U}\,ds_5^2.
 \ee
 Here we rescale a five-dimensional metric for later convenience.
The four-form flux $G$ can be dualized to a real scalar $f$\footnote{Hodge dual is taken with respect to the rescaled five-dimensional metric.} 
 \be
 *_5 G = -6\, e^{-12U}\, df.
 \ee
 By dimensional reduction on AdS$_6$ space, we obtain a five-dimensional effective Lagrangian as
 \be
 {\cal{L}}= \sqrt{g_5} \left( R-18\, \left(\partial\,U\right)^2+18\,e^{-12U}\left(\partial\,f\right)^2-30\, e^{-6U}\right).
 \ee
In two-dimensional target space, we have found three Killing vectors 
 \begin{align}
   K^1 &= \dfrac{\sqrt{2}}{6} \left( \partial_U +6\, f\, \partial_f \right),\nn\\
   K^2 &= f\, \partial_U +\left(\dfrac{1}{12}\,e^{12U}+3\, f^2 \right)\partial_f ,\nn\\
   K^3 &= -\dfrac{1}{3} \, \partial_f.
 \end{align}
 These Killing vectors generate $sl(2,\mathbb{R})$ algebra and correspond to a Cartan generator $h$, a positive root generator $e_1$ and a negative root generator $f_1$, respectively.
The coset representative in the Borel gauge is constructed by exponentiating a Cartan and a positive root generator as 
 \be
 {\cal{V}}= e^{-3 \sqrt{2}\, U\, h} e^{6\, f\, e_1}.
 \ee
This target space parametrize the coset $SL(2,\mathbb{R})/SO(1,1)$. 

The non-trivial scalar potential breaks  the global $SL(2,\mathbb{R})$ symmetry.
The Killing spinor equations are given by eq. (B.7) in \cite{Apruzzi:2014qva}.
With rescaling Dirac spinor $\eta$ as
 \be
 \eta \longrightarrow e^{-U}\, \eta
 \ee
and the following combinations of the Killing spinor equations\footnote{We call the Killing spinor equations (B.7) in \cite{Apruzzi:2014qva} as $\delta \tilde\chi_{\mp}$ and $\delta \tilde\psi_{\mu \pm}$}
\begin{align}
\delta \psi_{\mu } \equiv
 \left( \begin{array}{c} 
        \delta \tilde{\psi}_{\mu +}+ e^{-2U}\, \gamma_\mu\,  \delta \tilde{\chi}_{-}\\
        \delta \tilde{\psi}_{\mu -}- e^{-2U}\, \gamma_\mu\,  \delta \tilde{\chi}_{+}
 \end{array}\right), \qquad
\delta \chi \equiv  i\, 6\, e^{-2U} 
 \left( \begin{array}{c} 
      \delta \tilde{\chi}_{ +} \\ \delta \tilde{\chi}_{-}
 \end{array}\right),
\end{align}
the Killing spinor equations can be written covariantly
\be \label{KSE-11dim}
\delta \psi_{\mu } = \left(D_\mu + S_i\, \Gamma^i\, \gamma_\mu \right)\eta, \qquad
\delta \chi_ i = \left(P_{\mu i j}\, \gamma^\mu\, \Gamma^j + T_j\, \Gamma^j\, \Gamma_i
\right) \eta,
\ee
where
\begin{align}
D_\mu &=\nabla_\mu +\dfrac{1}{4} Q_{\mu i j}\, \Gamma^{ij}, \qquad
\eta=
 \left(\begin{array}{c}
   \eta_+ \\ \eta_-
 \end{array} \right), \qquad
 \Gamma^i = \left(\tau^2, -i\, \tau^1 \right).
\end{align}
Here two vectors $S_i$ and $T_i$ are introduced as
\be\label{solST-par}
 S_i = i\, e^{-3U} \left(1, \,0\right), \qquad  T_i = i\,3\, e^{-3U} \left(1, \,0\right).
\ee

The integrability conditions are examined with the general vectors $S_i$ and $T_i$. The gravitino-gravitino integrability condition is
\begin{align}\label{11d-gg}
&\gamma_\mu^{\phantom{\mu} \nu\rho} \left[{\mathcal D}_\nu , {\mathcal D}_\rho  \right]\eta \nn\\
&=\dfrac{1}{2} \left(P^{\phantom{\nu} i \phantom{j}}_{\nu \phantom{i} j}\,\gamma^\nu \,\Gamma^j -T_j\, \Gamma^i\, \Gamma^j\right) \gamma_\mu\, \delta \chi_i\nn\\
&+\left(R_{\mu\nu}-\dfrac{1}{2} R\, g_{\mu\nu}
    -\left( P_{\mu ij}\, P_{\nu}^{\phantom{\nu} ij}-\dfrac{1}{2}\, g_{\mu\nu}\, P_{\rho ij}\, P^{\rho ij}\right) 
    +g_{\mu\nu}\left(T_i \,T^i
    -24\, S_i\, S^i\right) \right) \gamma^\nu\, \eta \nn\\
&
-6 \left(\partial_\nu S_i +Q_{\nu i }^{\phantom{\nu i }j}\,S_j+\dfrac{1}{3}P_{\nu i }^{\phantom{\nu i }j}\,T_j  \right) \gamma^{\nu}_{\phantom{\nu}\mu}\, \Gamma^i\, \eta.
\end{align}
where ${\mathcal D}_\mu \equiv D_\mu + S_i\, \Gamma^i\, \gamma_\mu$. 
The gravitino-dilatino integrability condition is
\begin{align}\label{11d-gd}
\gamma^\mu \, D_\mu\, \delta \chi_i
&=P_{\nu i j }\, \gamma^\mu \, \gamma^\nu\, \Gamma^j\, \delta\psi_\mu
+T_j\, \Gamma^j\, \Gamma_i\, \gamma^\mu\, \delta\psi_\mu
+3\,\Gamma_i\,S^j\,\delta \chi_j - 3\,\Gamma_j\,S^j\,\delta \chi_i\nn\\
&+ \left(D_\mu \, P^\mu_{\phantom{\mu}ij}+ 5\, T_k\,S^k\,\eta_{ij}
-5\, T_i\,S_j-5\, T_j\,S_i \right) \Gamma^j\, \eta\nn\\
&-T^j\,S^k\,\Gamma_{ijk}
+\left(\partial_\mu\, T_i + Q_{\mu i}^{\phantom{\mu i}j}\,T_j+ 3\,P_{\mu i}^{\phantom{\mu i}j}\,S_j\right) \gamma^\mu\,\eta\nn\\
&+\left(\partial_\mu\, T_j+ Q_{\mu j}^{\phantom{\mu j}k}\,T_k+ 3\,P_{\mu j}^{\phantom{\mu j}k}\,S_k\right)\Gamma^{j}_{\phantom{j}i}\, \gamma^\mu\,\eta.
\end{align}
In both integrability conditions, the first lines vanish by Killing spinor equations $\delta\psi_\mu=\delta \chi_i=0$. The second lines give the Einstein equation and the scalar equations of motion, respectively. Finally the rest give the equations for $S_i$ and $T_i$, which should be satisfied to yield conventional integrability conditions.
We find a solution to these equations as
\be\label{solST}
T_i=3S_i =\left(t,0\right), \qquad t=\alpha\, {\cal{V}}_{11},
\ee
where $\alpha$ is an integration constant.

As in the case of AdS$_6$ in IIB supergravity, we generalize this construction to larger coset space $SL(n,\mathbb{R})/SO(p,q)$. 
First, we have to modify the dilatino variation \eqref{KSE-11dim} as
\be
\delta \chi_ i= \left(P_{\mu i j}\, \gamma^\mu\, \Gamma^j + T_j\, \Gamma^j\, \Gamma_i +\dfrac{n-2}{n} T_j\, \Gamma_i\, \Gamma^j\right) \eta,
\ee
to satisfy $\Gamma^i\, \delta \chi_i=0$ condition.
Here $i=1,\cdots,n$ and $T_i$ is a n-component vector.
After a tedious calculation, the gravitino-gravitino integrability condition reduces to
\begin{align}
&\gamma_\mu^{\phantom{\mu} \nu\rho} \left[{\mathcal D}_\nu , {\mathcal D}_\rho  \right]\eta \nn\\
&= \left( \text{r.h.s of}\,  \eqref{11d-gg}\right)
-\dfrac{n-2}{2n} T_j\, \Gamma^j\, \Gamma^i \gamma_\mu\, \delta \chi_i
-g_{\mu\nu}\dfrac{(n-2)^2}{2n} T_i\, T^i\, \gamma^\nu\,\eta.
\end{align}
There are two additional contributions to \eqref{11d-gg}. The first term vanishes due to Killing spinor equations. The second term give the extra contribution to the scalar potential in the Einstein equation. As a result, we have the Einstein equation 
\be
R_{\mu\nu}-\dfrac{1}{2} R\, g_{\mu\nu}
    -\left( P_{\mu ij}\, P_{\nu}^{\phantom{\nu} ij}-\dfrac{1}{2}\, g_{\mu\nu}\, P_{\rho ij}\, P^{\rho ij}\right) 
    +g_{\mu\nu}\left( \dfrac{n}{2}T_i \,T^i
    -\dfrac{(n-2)^2}{2n}T_i\,T^i
    -24\, S_i\, S^i\right)=0,
\ee
where the scalar potential is
\be\label{pot-mth}
V=n\,T_i \,T^i
    -\dfrac{(n-2)^2}{n}T_i\,T^i
    -48\, S_i\, S^i.
\ee
The gravitino-dilatino integrability condition is more involved.
\begin{align}
\gamma^\mu \, D_\mu\, \delta \chi_i
&=\left( \text{r.h.s of}\,  \eqref{11d-gd}\right)\nn\\
&+\dfrac{n-2}{n}\Big\{\gamma^\mu \,T_j\, \Gamma_i\, \Gamma^j\, \delta\psi_\mu
 -3\Gamma_i\, S^j\, \delta\chi_j
 +\left(\partial_\mu\, T_j+ Q_{\mu j}^{\phantom{\mu j}k}\,T_k+3 P_{\mu j}^{\phantom{\mu j}k}\,S_k \right)\,\gamma^\mu\,\Gamma_i\,\Gamma^j\,\eta \nn\\
&\phantom{\dfrac{n-2}{n}\Big\{\gamma^\mu }
+ 4\,T^j\,S^k\,\Gamma_{ijk} 
+\left(4\,T_i\,S_j +2\,T_j\,S_i-8\,T_k\,S^k\,\eta_{ij}  \right)\Gamma^j\,\eta\Big\}\nn\\
&+3 \left(\dfrac{n-2}{n}\right)^2\Big\{ -T^j\,S^k\,\Gamma_{ijk}+
\left(-T_i\,S_j +T_j\,S_i + T_k\,S^k\,\eta_{ij}  \right)\Gamma^j\,\eta\Big\}.
\end{align}
Similarly, the only non-trivial contribution appears in the scalar equations of motion as
\begin{gather}
D_\mu \, P^\mu_{\phantom{\mu}ij}+ 5\, T_k\,S^k\,\eta_{ij}-5\, T_i\,S_j-5\, T_j\,S_i\, \nn\\
+\dfrac{n-2}{n}\left(4\,T_i\,S_j +2\,T_j\,S_i-8\,T_k\,S^k\,\eta_{ij}  \right) 
+3 \left(\dfrac{n-2}{n}\right)^2 \left(-T_i\,S_j +T_j\,S_i + T_k\,S^k\,\eta_{ij}  \right)=0.
\end{gather}
Even though we have examined the integrability conditions with the modified dilatino variation, the equations for $S_i$ and $T_i$, which should be satisfied and eliminated in the integrability conditions, are not changed except they are now n-component equations. They are easily solved by
\be\label{solST-gen}
T_i=3S_i =\left(t,0,\cdots,0\right), \qquad t=\alpha\, {\cal{V}}_{11},
\ee
which is the most simple generalization of \eqref{solST}.
The scalar potential is determined by $(1,1)$-component of the coset representative $\cal{V}$.

\bibliography{ref}{}
\end{document}